\documentclass[10pt,a4paper]{article}
\usepackage{jheppub_kim}
\usepackage{pdflscape}
\usepackage{amsmath}
\usepackage{amssymb}
\usepackage{dcolumn}
\usepackage{bm}
\usepackage{color}
\usepackage{epsfig}
\usepackage{amsfonts}
\usepackage{graphicx}
\usepackage{subfigure}
\usepackage{dcolumn}

\begin{document}
\title{ Non-Pertubative Quantum Corrections to a Born-Infeld Black Hole and its Information Geometry}
\author[a]{Behnam Pourhassan,}
\author[b]{Mohsen Dehghani,}
\author[c]{Mir Faizal,}
\author[d]{Sanjib Dey}

\affiliation[a] {School of Physics, Damghan University, Damghan, 3671641167, Iran.}
\affiliation[b] {Department of Physics, Razi University, Kermanshah, Iran}
\affiliation[c] {Department of Physics and Astronomy, University of Lethbridge, Lethbridge, Alberta, T1K 3M4, Canada.}
\affiliation[c] {Irving K. Barber School of Arts and Sciences, University of British Columbia, Kelowna, British Columbia, V1V 1V7, Canada.}
\affiliation[c] {Canadian Quantum Research Center 204-3002 32 Ave Vernon, BC V1T 2L7 Canada.}
\affiliation[d] {Department of Physical Sciences, Indian Institute of Science Education and Research Mohali, Sector 81, SAS Nagar, Manauli 140306, India.}

\emailAdd{b.pourhassan@du.ac.ir}
\emailAdd{m.dehghani@razi.ac.ir}
\emailAdd{mirfaizalmir@googlemail.com}
\emailAdd{sanjibdey4@gmail.com}

\abstract{ We study the non-perturbative quantum corrections to  a Born-Infeld black hole in a spherical cavity. These  quantum  corrections produce a non-trivial short distances modification to the relation between the entropy and area of this black hole. The non-perturbative quantum correction appears as an exponential term in the black hole entropy. This in turn modifies the thermodynamics of a given system, for example reduced value of the Helmholtz free energy. Moreover, the first law of black hole thermodynamics modified due to quantum corrections. We also investigate the effect of such non-perturbative corrections on the information geometry of this system. This is done using some famous information metrics.}


\maketitle

\section{Introduction}

It is known that black holes behave as hot objects, and have an entropy equal to a quarter of their area. In fact, this entropy of a black hole is the maximum entropy that can be associated with any object of the same volume \cite{1a, 2a}. Thus, the maximum entropy that a region of space can contain scales with its area rather than its volume, and this observation has led to the development of the holographic principle \cite{4a, 5a}. However, it has also been argued that the holographic principle could get modified at Planck scale, due to the modification to the structure of spacetime by quantum fluctuations \cite{6a, 7a}. Thus, we expect that the relation between the entropy and area would also get modified by quantum corrections to the structure of spacetime.

Such quantum corrections to the entropy of a black hole have been studied using various different approaches, and it has been demonstrated that the quantum corrections even modify other aspects of black holes at short distances.
In fact, AdS/CFT has been used to analyze the quantum correction to the entropy of various AdS black holes \cite{18, 18a, 18b, 18c, 18d}. Such quantum correction to the entropy has also been calculated using the extremal limit of a  black hole \cite{19, 19a}. Non-perturbative quantum correction to the entropy of a black hole has also been obtained using the density of microstates associated with conformal blocks \cite{Ashtekar}. Such quantum corrections have also been obtained from the Cardy formula for black holes \cite{Govindarajan}. It is also possible to use the Rademacher expansion to obtain corrections to the entropy of a black hole \cite{29}. Thus, it is well established that the entropy of a black hole would get corrected at short distances due to quantum fluctuations. It has been argued in Jacobson formalism that the geometry of space-time emerges from thermodynamics, as an emergent theory \cite{gr12}. So, using the Jacobson formalism, it can be demonstrated that the thermal fluctuations in the thermodynamics of black holes produce quantum fluctuations in the geometry of space-time \cite{gr14}. The thermal fluctuations to the thermodynamics of various black holes have been thoroughly studied \cite{32, 32a, 32b, 32c, 32d}. In fact, as in the Jacobson formalism, these thermal fluctuations are related to quantum fluctuations \cite{gr12, gr14}. Quantum corrections to the entropy of various black holes have also been studied \cite{40a, 40b, 40c, 40d}.

As these corrections occur due to quantum fluctuations, they can be neglected for large black objects. This is because large black holes have a small temperature, and so the thermal fluctuations can be neglected for large black holes. As in the Jacobson formalism \cite{gr12, gr14}, these thermal fluctuations produce the quantum fluctuations in the geometry of space-time, we can also neglect quantum fluctuations in the geometry of space-time for large black holes. However, as these black holes reduce in size due to the Hawking radiation, we cannot neglect the effects of quantum fluctuations on them.
Thus, we need to consider the leading order quantum corrections to the entropy of such black holes to analyze the short distance corrections to the thermodynamics \cite{40a, 40b, 40c, 40d}. However, as the black holes reduce further in size, we need to consider the full non-perturbative corrections (not just leading order) to the entropy of a black hole. It has been recently proposed that such corrections are represented by an exponential function of the original entropy, and they become important for black holes at short distances \cite{2007.15401}. It has also been demonstrated that such non-perturbative corrections to the entropy of a black hole can be obtained using the Kloosterman sums \cite{Dabholkar}. This is done using AdS/CFT correspondence for massless supergravity fields near the horizon \cite{ds12, ds14}.

As these non-perturbative corrections \cite{2007.15401, Dabholkar} can be motivated from string theoretical effects \cite{ds12, ds14}, it is important to analyze their effects on the geometries motivated by string theory. It is known that the action for D-branes can be described by a Born-Infeld action \cite{bi12, bi14}. This has motivated the construction of Born-Infeld black holes \cite{bibh12, bibh14}. It may be noted that the field theory dual to such Born-Infeld black holes has also been investigated using the AdS/CFT correspondence \cite{biab, bicd}. It has been demonstrated that the non-linear Born-Infeld in the action produces non-trivial modifications to the thermodynamics of these Born-Infeld black holes \cite{bi15, bi16, bi17, bi18}. It is possible to place a black hole inside a cavity \cite{ca12, ca14}. In fact, it is possible to obtain interesting results for a system of D-branes by placing it in a cavity \cite{ca15, ca16, da12, da14}. As Born-Infeld action is motivated from D-branes \cite{bi12, bi14}, the properties of  Born-Infeld black holes have also been investigated by placing them in a cavity \cite{m1, m2}. It has been demonstrated that the Born-Infeld black holes in a cavity can have interesting thermodynamic behavior  \cite{JHEP, JHEP1}. So, in this paper, we will study the effects of such non-perturbative corrections on the thermodynamics of a Born-Infeld black hole in a cavity. We will also analyze the non-perturbative corrections to the information-theoretical geometry of this system. This will be done using different informational theoretical metrics for this system \cite{q1, q2, w1, w2, r1, r2, HPEM}. It will be observed that these non-perturbative corrections will produce a non-trivial modification to the information-theoretical geometry of this system.

\section{Born-Infeld Black Hole in a Cavity}\label{sec2}
Let us first briefly review a Born-Infeld black hole in a cavity \cite{m1, m2}. A Born-Infeld black hole inside a four-dimensional space-time manifold $\mathcal{M}$, with a time-like boundary $\partial\mathcal{M}$ at $r=r_{B}$ can be constructed by coupling gravity to a Born-Infeld electromagnetic field $A_{\mu}$ with the field strength tensor $F_{\mu\nu}=\partial_{\mu}A_{\nu}-\partial_{\nu}A_{\mu}$. Thus, we can write the action for this system as \cite{JHEP}
\begin{eqnarray}\label{action}
\mathcal{S}=&&\int_{\mathcal{M}}{d^{4}x\sqrt{-g}\left[R+\frac{1}{a}\left(1-\sqrt{1+\frac{a}{2}F^{\mu\nu}F_{\mu\nu}}\right)\right]}\nonumber\\
&&-2\int_{\partial\mathcal{M}}{d^{3}x\sqrt{-\gamma}\left[(K-K_{0})+\frac{A_{\mu}n_{\nu}F^{\mu\nu}}{\sqrt{1+\frac{a}{4}F^{\mu\nu}F_{\mu\nu}}}\right]},
\end{eqnarray}
in unit of $16\pi G$, with the last term being the surface term. This solution can be obtained from string theory, and in that case, the coupling parameter $a$ is related to the string slope parameter $\alpha^\prime$ as $a=(2\pi\alpha^{\prime})^2$. Here, $K$ and $K_0$ represent the extrinsic curvature and a subtraction term, respectively. The subtraction term $K_0$ is placed in the surface term to make sure that the Gibbons-Hawking-York term vanish. Furthermore, $\gamma$ is the metric on the boundary and $n_{\nu}$ is the unit normal of the surface vector pointing in the outward direction.

\begin{figure}[h!]
\begin{center}$
\begin{array}{cccc}
\includegraphics[width=50 mm]{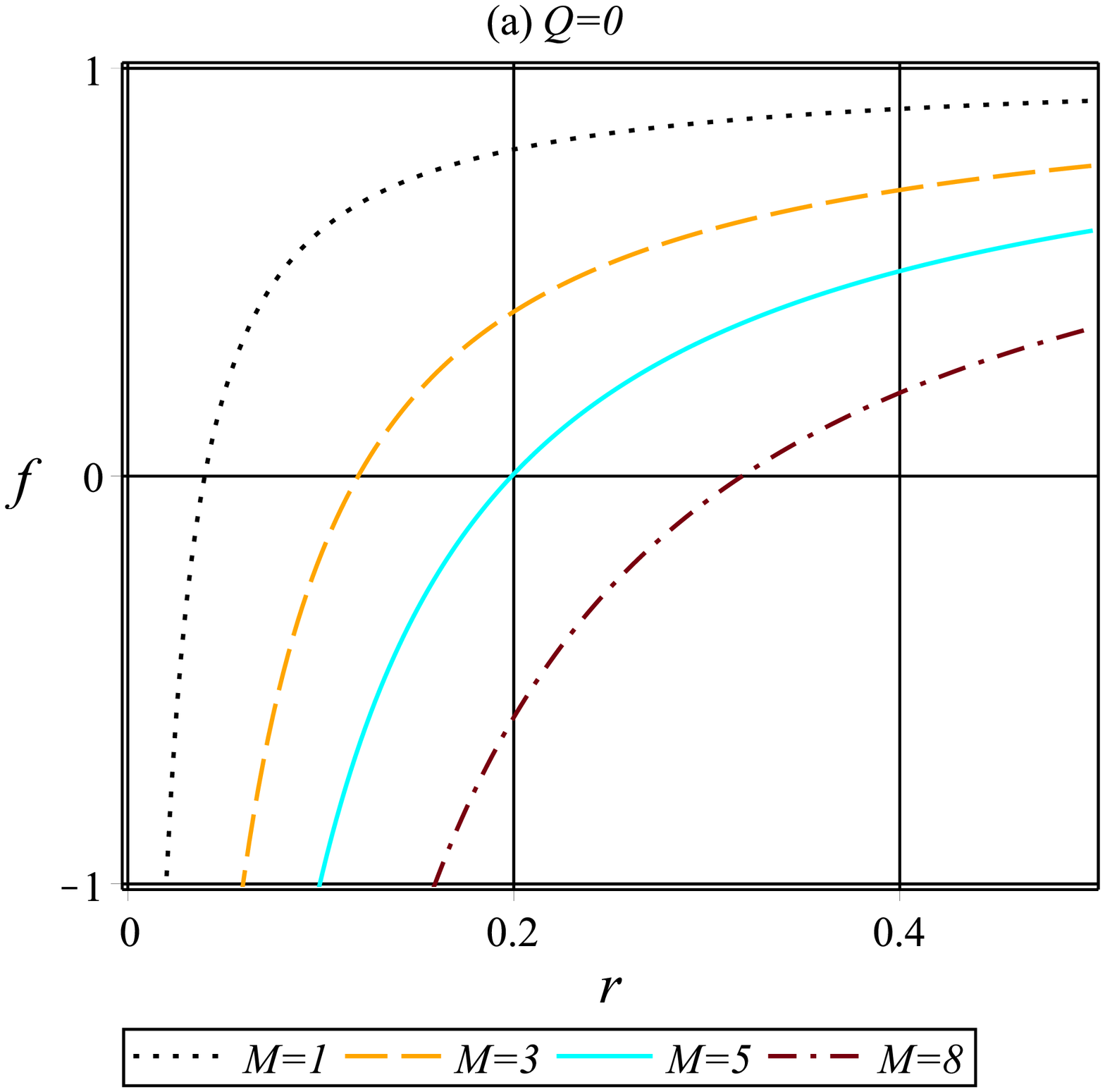}\includegraphics[width=50 mm]{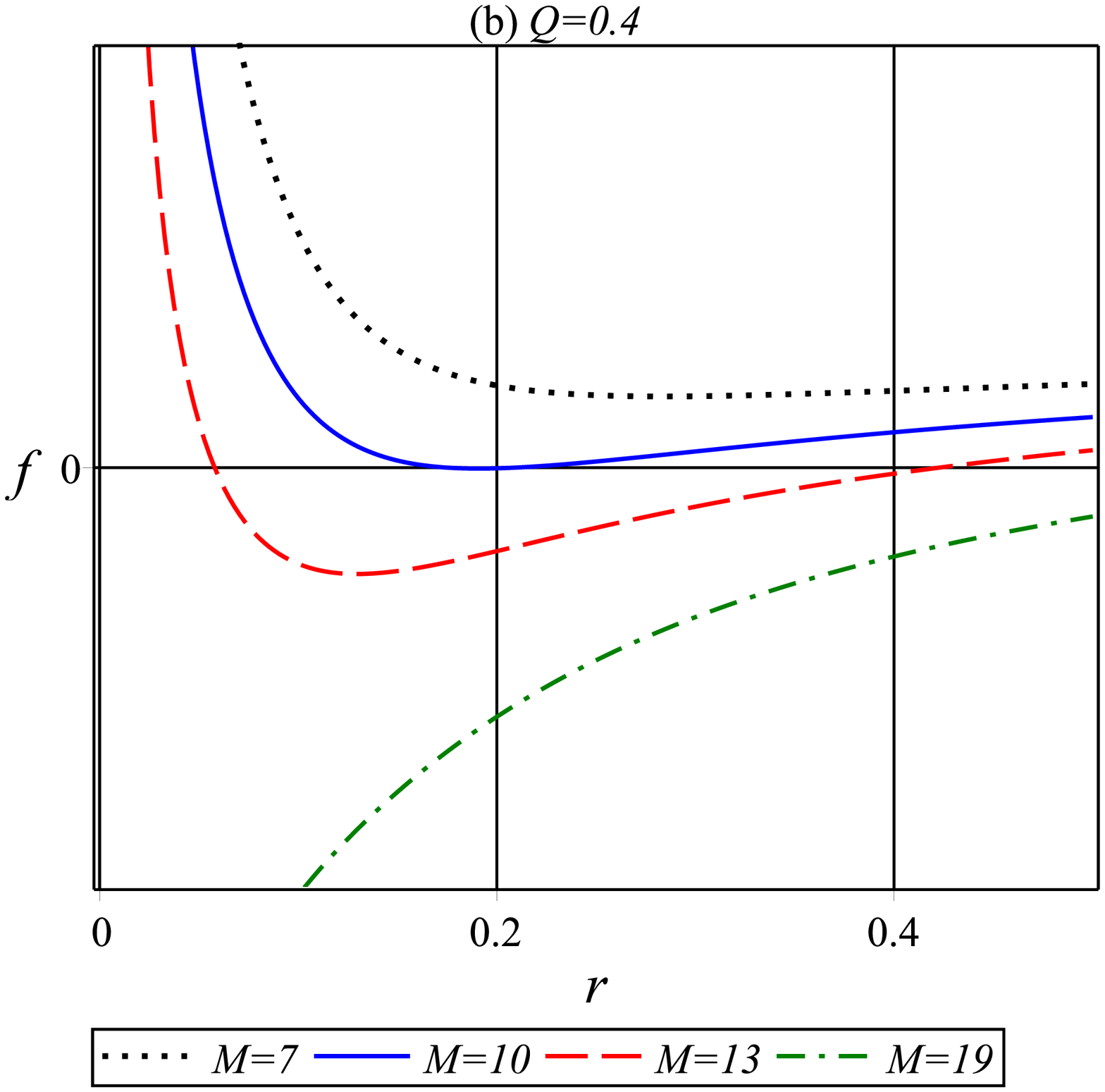}\includegraphics[width=50 mm]{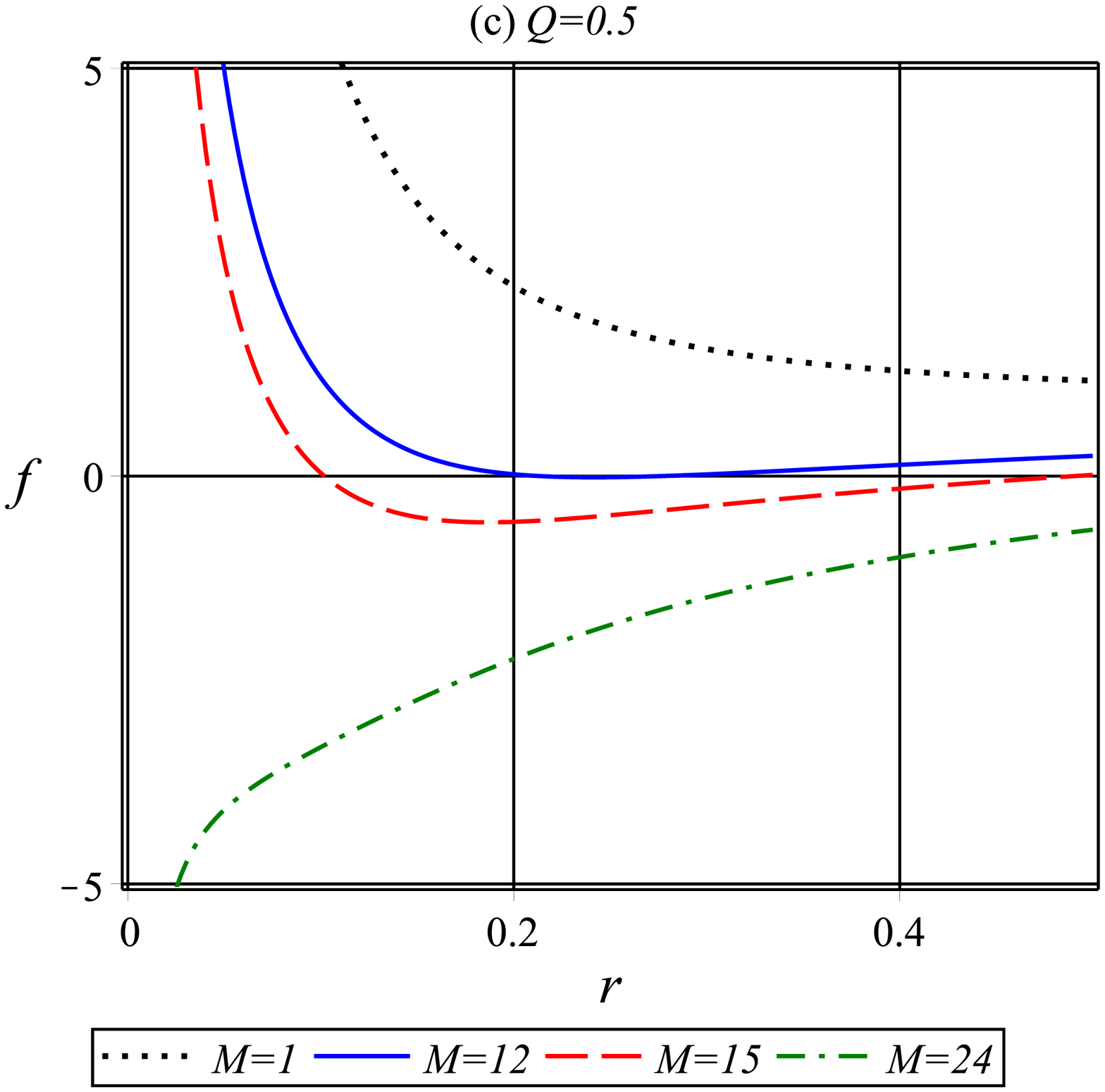}
\end{array}$
\end{center}
\caption{Horizon structure of Born-Infeld black hole in a cavity for $a=\frac{1}{300}$.}
\label{fig1}
\end{figure}

The static spherically symmetric black hole (of mass $M$ and charge $Q$) solution arising from the action (\ref{action}) is given by \cite{m1, m2}
\begin{equation}\label{metric}
ds^{2}=-f(r)dt^{2}+\frac{dr^{2}}{f(r)}+r^{2}d\Omega_{2}^{2},
\end{equation}
with
\begin{equation}\label{metric f}
f(r)=1-\frac{M}{8\pi r}-\frac{Q^{2}}{6\sqrt{r^{4}+aQ^{2}}+6r^{2}}+\frac{Q^{2}}{3r^{2}}\hspace{1mm} {}_{2}F_{1}(\frac{1}{4},\frac{1}{2},\frac{5}{4};-\frac{aQ^{2}}{r^{4}}),
\end{equation}
where $d\Omega_{2}^{2}=d\theta^{2}+\sin^{2}\theta d\phi^{2}$, and ${}_{2}F_{1}$ is generalized hypergeometric function. In the limit $Q\rightarrow0$, the above solution reproduces the Schwarzschild black hole, which is thermodynamically unstable. In Fig. \ref{fig1}, we demonstrate the horizon structure of the charged back hole (\ref{metric}). The black dotted line in this figure shows a naked singularity for a specific choice of $M$ ($M=7$). A relatively larger value of $M$ yields the extremal case (see, the solid blue line), where both the horizons coincide. By increasing $M$ further one can obtain both the inner ($r_{-}$) and outer ($r_{+}$) horizons, as shown by the red dashed line. For example, by choosing $Q=0.4$, $a=0.003$ and $M=13$, we have $r_{+}\approx0.4$. Furthermore, larger values of $M$ (dash-dotted green line) yields only one horizon.

\begin{figure}[h!]
\begin{center}$
\begin{array}{cccc}
\includegraphics[width=70 mm]{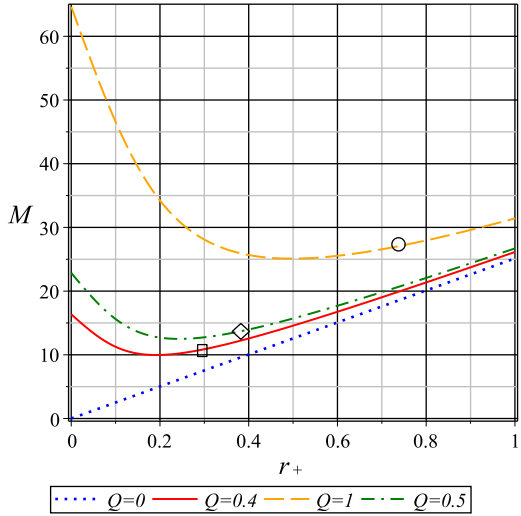}
\end{array}$
\end{center}
\caption{Mass of Born-Infeld black hole in a cavity for $a=\frac{1}{300}$.}
\label{fig2}
\end{figure}

The mass of the Born-Infeld black hole (\ref{metric}) for $f(r)=0$ is given by
\begin{equation}\label{M}
M=8\pi r_{+}\left[1-\frac{Q^{2}}{6\sqrt{r_{+}^{4}+aQ^{2}}+6r_{+}^{2}}+\frac{Q^{2}}{3r_{+}^{2}}\hspace{1mm} {}_{2}F_{1}(\frac{1}{4},\frac{1}{2},\frac{5}{4};-\frac{aQ^{2}}{r_{+}^{4}})\right].
\end{equation}
The variation of the mass (\ref{M}) with the horizon radius $r_{+}$ is plotted in Fig. \ref{fig2} for different values of the charge $Q$. Eq. (\ref{M}) indicates that the mass $M$ varies linearly with $r_{+}$ in the limit $Q\rightarrow0$. In the case of a finite electric charge, the mass of the black hole decreases with the decrease of $r_{+}$ and reaches a minima at a critical value ($r_{c}$). For example, for $Q=0.4$, $r_{c}\approx0.2$ (see the minimum of red solid line of Fig. \ref{fig2}). By decreasing the horizon radius $r_{+}$ further, the mass of the black hole increases, which is in agreement with its behavior described in \cite{m1}. It is interesting to note that the black hole mass does not become zero when $r_{+}$ vanishes. There are some specific points with circle and square which will explain later (end of this section). It is possible for the horizon radius $r_{+}$ to vanish when the mass is non-zero. In this paper, we shall analyze this case by studying the behavior of this black hole at a small horizon radius. It has been argued that at such scales the black hole entropy should get corrected by an exponential term \cite{2007.15401}.

In the next section, we show the origin of such a correction and discuss about the modification of the thermodynamics due to the exponentially corrected entropy.

\section{Quantum Correction}\label{sec3}
\begin{figure}[h!]
\begin{center}$
\begin{array}{cccc}
\includegraphics[width=60 mm]{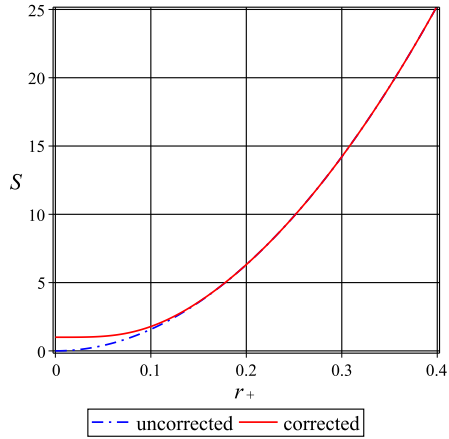}
\end{array}$
\end{center}
\caption{Entropy of Born-Infeld black hole in a cavity in terms of horizon radius.}
\label{fig2.5}
\end{figure}

It is expected that the quantum corrections would correct the relation between area and entropy of a black hole \cite{40a, 40b, 40c, 40d}. In fact, such corrections have been studied mostly perturbatively \cite{log0, P1, higher3}. However, it is important to analyze the full non-perturbative quantum corrections to this black hole. Now, as the quantum corrections to the geometry of space-time can be obtained from thermal fluctuations in the Jacobson formalism \cite{gr12, gr14}, we would like to analyze such non-perturbative corrections to the area entropy relation of Born-Infeld black hole
\begin{equation}\label{S0}
S_{0}=16\pi^{2}r_{+}^{2}.
\end{equation}
It may be noted that perturbative quantum correction in the form of logarithmic correction \cite{log1, log2}, and even higher-order perturbative corrections \cite{higher1, higher2} have been studied using thermal fluctuations. Here will analyze non-perturbative exponential corrections to the entropy of this system using thermal fluctuations.  Now in statistical mechanics, the entropy of a given system is related to the number of measurable microstates
\begin{equation}\label{SO}
S_{0}=\ln{\Omega},
\end{equation}
where we have used the units of the Boltzmann constant ($k_{B}=1$).
Now if the microstates of this black hole can be represented by a conformal field theory, then it is known that the conformal field theory corrects represents the entropy and its thermal fluctuations for such a black hole \cite{Govindarajan, 29}. Now, these perturbative thermal fluctuations seem to have a universal form, and this universal form would imply that the corrections to the entropy would also have a universal form \cite{32, 32a, 32b, 32c, 32d}
\begin{equation}
S_{per} = \ln{\Omega} + f(\Omega) = S_0 + \frac{1}{2} \ln (T^2 S_0) + ...
\end{equation}
It has also been argued that the non-perturbative corrections to a black hole will be expressed by an exponential function of the original entropy \cite{2007.15401, Dabholkar}. Now these corrections should also be produced by a suitable function of $\Omega$, and so they can be written as
\begin{equation}\label{Sm}
S_{m}=\frac{1}{\Omega}=e^{-S_{0}}.
\end{equation}
As the perturbative corrections are universal, we propose that this non-perturbative correction is also universal (as both are obtained as functions of $\Omega$). Hence, we can write the total non-perturbative quantum corrected entropy of Born-Infeld black hole as
\begin{equation}\label{S}
S=S_{0}+S_{m}=16\pi^{2}r_{+}^{2}+e^{-16\pi^{2}r_{+}^{2}}.
\end{equation}
The exponential term is negligible for a large black hole, however, it becomes relevant when the area of the black hole is small. It is illustrated by Fig. \ref{fig2.5}. It may be noted that this correction can produce an important quantum effect for black holes at short distances. \\

\begin{figure}[h!]
\begin{center}$
\begin{array}{cccc}
\includegraphics[width=70 mm]{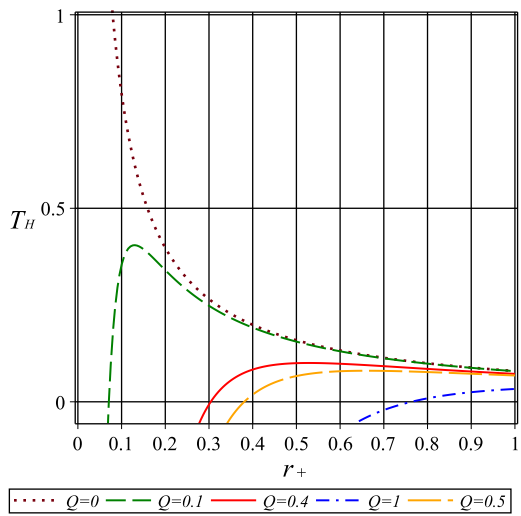}
\end{array}$
\end{center}
\caption{Hawking temperature of Born-Infeld black hole in a cavity for $a=\frac{1}{300}$.}
\label{fig3}
\end{figure}

The Hawking temperature of black holes is obtained by using the relations (\ref{metric f}) and (\ref{M}) as follows
\begin{eqnarray}\label{tempH}
T_{H}=\frac{1}{4\pi}\left(\frac{df(r)}{dr}\right)_{r=r_{+}}=&&\frac{aQ^{4}}{30\pi r_{+}^{7}}\hspace{1mm} {}_{2}F_{1}(\frac{5}{4},\frac{3}{2},\frac{9}{4};-\frac{aQ^{2}}{r_{+}^{4}})
-\frac{Q^{2}}{6\pi r_{+}^{3}}\hspace{1mm}{}_{2}F_{1}(\frac{1}{4},\frac{1}{2},\frac{5}{4};-\frac{aQ^{2}}{r_{+}^{4}})\nonumber \\
&& -\frac{(\sqrt{r_{+}^{4}+aQ^{2}}-2r_{+}^{2}-6a)Q^{2}-6r_{+}^{2}(r_{+}^{2}+\sqrt{r_{+}^{4}+aQ^{2}})}{24\pi r_{+}(r_{+}^{2}+\sqrt{r_{+}^{4}+aQ^{2}})\sqrt{r_{+}^{4}+aQ^{2}}}.
\end{eqnarray}
In Fig. \ref{fig3}, we show the behavior of Hawking temperature with the horizon radius. Now when $Q\neq0$, we can see there is a minimum value of the horizon radius ($r_{m}$), where the Hawking temperature vanishes. Previously, in Fig. \ref{fig2}, we showed the corresponding mass at this point, and they were depicted by a small square in the red solid line, and a circle in the green dashed line. Therefore, it is obvious that $r_{m}>r_{c}$, which means, before the black hole reaches to the minimum mass, the Hawking temperature vanishes. The specific heat of black holes also is zero at this point, hence, the Hawking radiation will stop at this point, and the black hole will not evaporate any further. So, it may create a black remnant of finite mass in zero temperature and zero specific heat, which may solve the information loss paradox of black holes \cite{43,44}. However, a more important point which was discussed earlier is that the black hole entropy should get corrected by an exponential term at this range of horizon radius, modifying thermodynamics of this system.

\section{ Corrected Thermodynamics}
\begin{figure}[h!]
 \begin{center}$
 \begin{array}{cccc}
\includegraphics[width=70 mm]{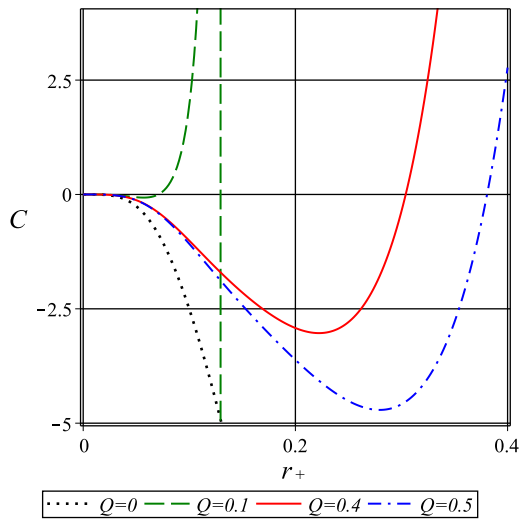}
 \end{array}$
 \end{center}
\caption{Specific heat of Born-Infeld black hole in a cavity for $a=\frac{1}{300}$.}
 \label{fig4}
\end{figure}

\begin{figure}[h!]
 \begin{center}$
 \begin{array}{cccc}
\includegraphics[width=70 mm]{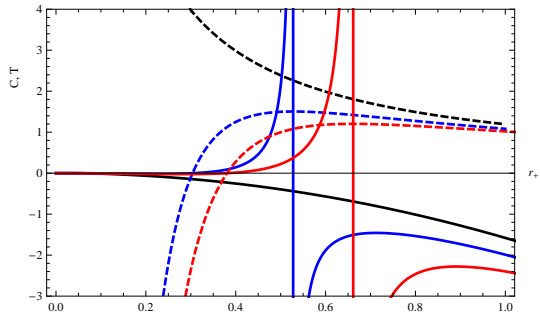}
 \end{array}$
 \end{center}
\caption{Specific heat (solid lines) and temperature (dashed lines) of Born-Infeld black hole in a cavity for $a=\frac{1}{300}$. $Q=0$ (black), $Q=0.4$ (blue), $Q=0.5$ (red).}
 \label{fig4.5}
\end{figure}

Now we can analyze the effects of these corrections on the thermodynamics of this system. Using the expressions of entropy in (\ref{S}) and temperature in (\ref{tempH}), one can study the specific heat using the following relation
\begin{equation}\label{C}
C=T_{H}\frac{d S}{dT_{H}}.
\end{equation}
In Fig. \ref{fig4}, we show the behavior of specific heat of Born-Infeld black hole in a cavity for different black hole charges. The special case of $Q=0$ is completely unstable, like a Schwarzschild black hole. However, for a charged black hole, we notice a first-order phase transition (unstable to stable). With the decrease of horizon radius, the black hole goes to the unstable phase. Indeed, for $r_{+}=r_{m}$, the specific heat become zero. So, a black remnant forms before the black hole enters the unstable phase.
In order to show phase transition points clearly, we combine Fig. \ref{fig3} and Fig. \ref{fig4} in Fig. \ref{fig4.5}. Curves of Fig. \ref{fig4.5} show that there is a point of first-order phase transition at the vanishing point of the black hole specific heat where the black hole temperature vanishes too. We label this point as $r_+=r_{ext}$. Also, there is a point of second-order phase transition identified by the divergent point of the specific heat, which we label by $r_+ =r_1$.  The black holes with the horizon radii in the range $ r_ {ext}< r_+<r_1$ are locally stable.\\

\begin{figure}[h!]
 \begin{center}$
 \begin{array}{cccc}
\includegraphics[width=60 mm]{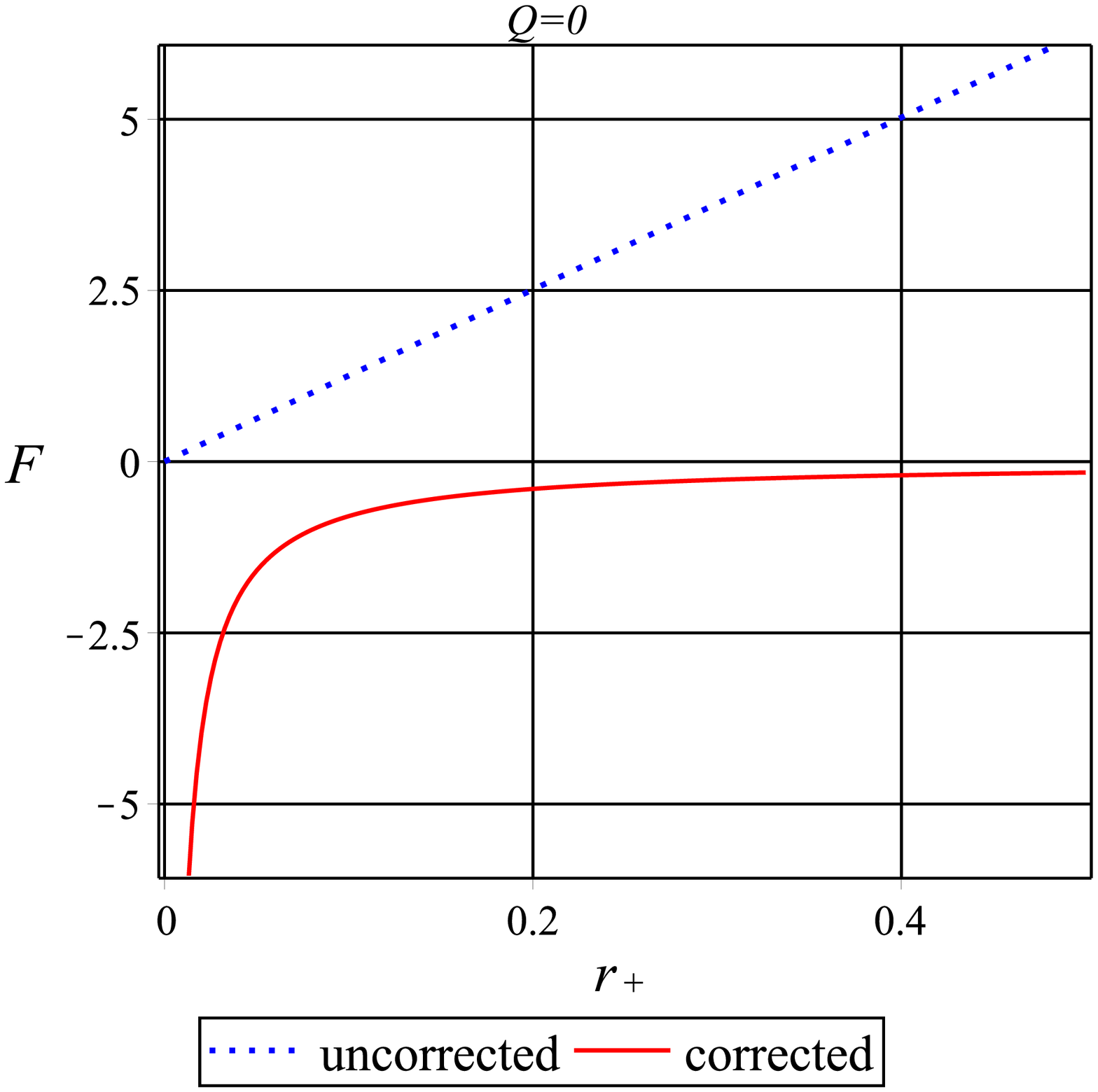}\includegraphics[width=60 mm]{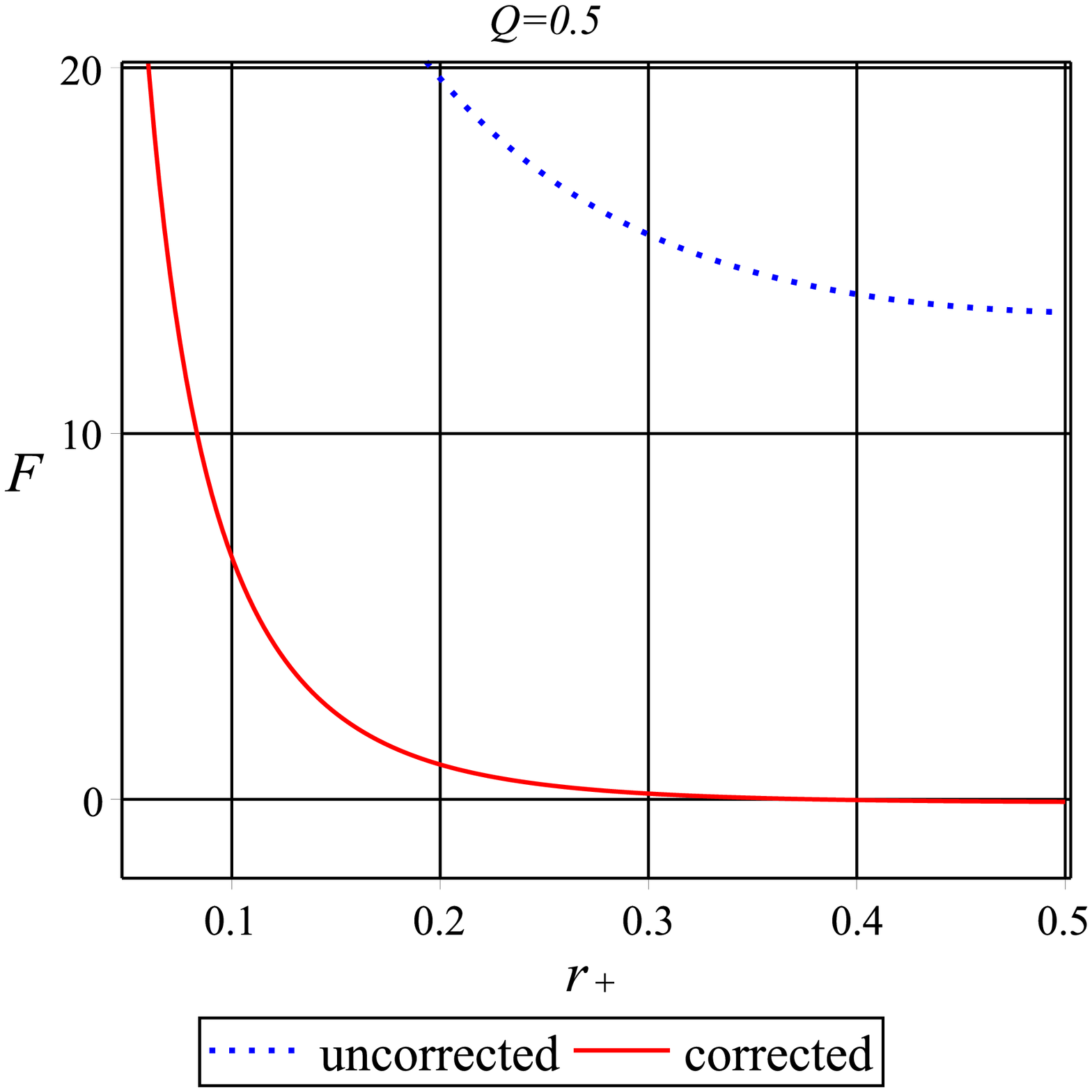}
 \end{array}$
 \end{center}
\caption{Helmholtz free energy of Born-Infeld black hole in a cavity for $a=\frac{1}{300}$.}
 \label{fig5}
\end{figure}

It is interesting to calculate the Helmholtz free energy at the point $r_{+}=r_{m}$. To the first order approximation ($e^{-16\pi^{2}r_{+}^{2}}\approx 1-16\pi^{2}r_{+}^{2}$), it becomes
\begin{eqnarray}\label{F0}
F=-\int{SdT_{H}}\approx &&\frac{Q^{2}}{30\pi r_{+}^{7}}\left[5r_{+}^{4}\hspace{1mm}{}_{2}F_{1}(\frac{1}{4},\frac{1}{2},\frac{5}{4};-\frac{aQ^{2}}{r_{+}^{4}})
-aQ^{2}\hspace{1mm} {}_{2}F_{1}(\frac{5}{4},\frac{3}{2},\frac{9}{4};-\frac{aQ^{2}}{r_{+}^{4}})\right]\nonumber\\
&&+\frac{3r_{+}^{4}+aQ^{2}-3(r_{+}^{2}+2a)\sqrt{r_{+}^{4}+aQ^{2}}}{24a\pi r_{+}\sqrt{r_{+}^{4}+aQ^{2}}}.
\end{eqnarray}
In Fig. \ref{fig5}, we plot the behavior of Helmholtz free energy $F$ of Born-Infeld black hole in a cavity, with the variation of horizon radius $r_+$. We denote the uncorrected Helmholtz free energy by $F_{0}$, where  the exponential term in (\ref{S}) is neglected. In both of the panels of Fig. \ref{fig5}, we can see that the overall effect of the non-perturbative  quantum correction is to reduce the Helmholtz free energy. The left panel ($Q=0$) is more interesting, where we notice that the Helmholtz free energy is positive and grows linearly with the increasing horizon radius, if we neglect the quantum correction. However, in the presence of quantum corrections, it is completely negative approaching zero at larger $r_{+}$. Also, for the charged case (right panel), the Helmholtz free energy is positive which becomes zero at large $r_{+}$. This means that the black remnant has a positive energy.\\

\begin{figure}[h!]
 \begin{center}$
 \begin{array}{cccc}
\includegraphics[width=50 mm]{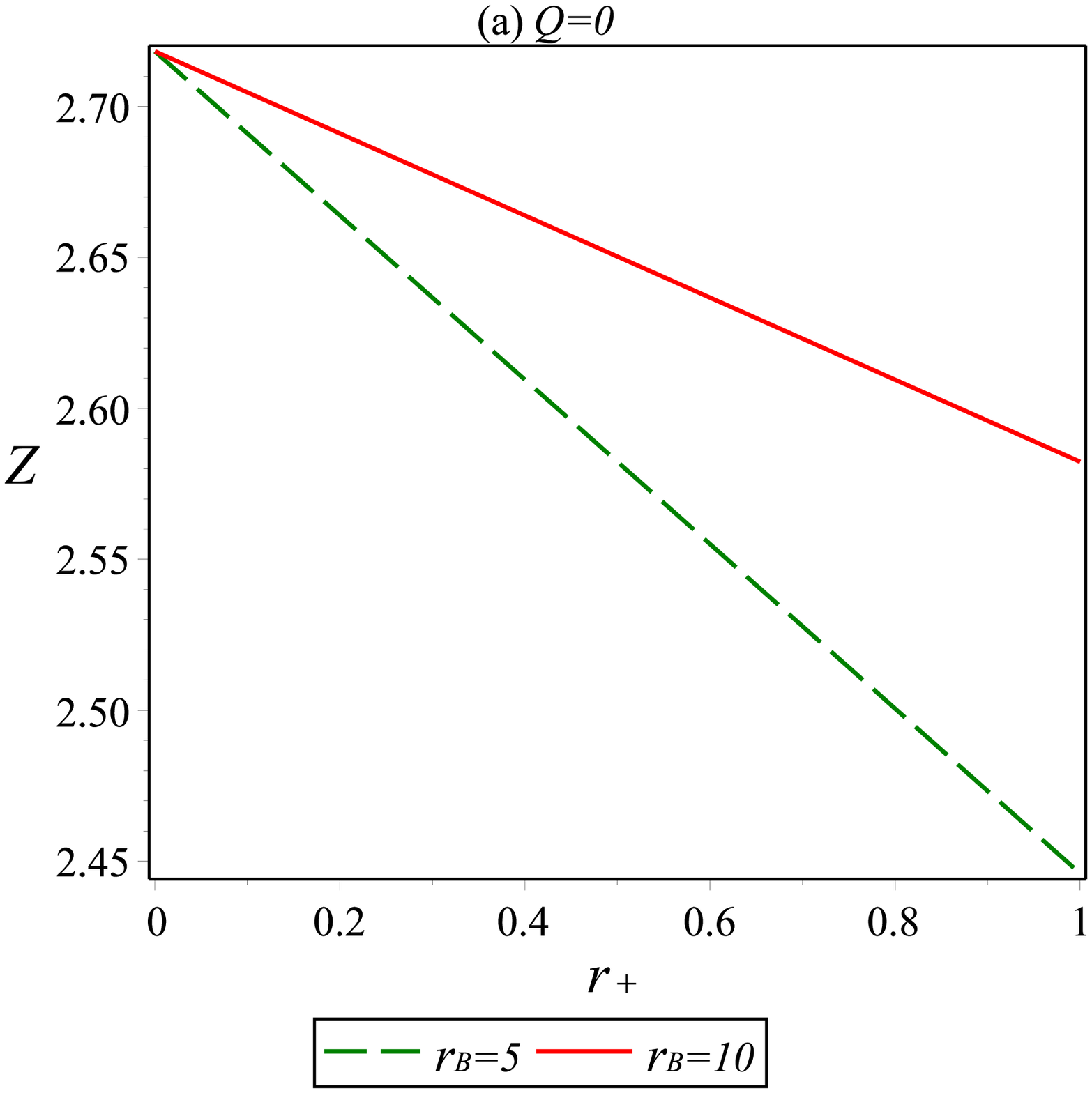}\includegraphics[width=50 mm]{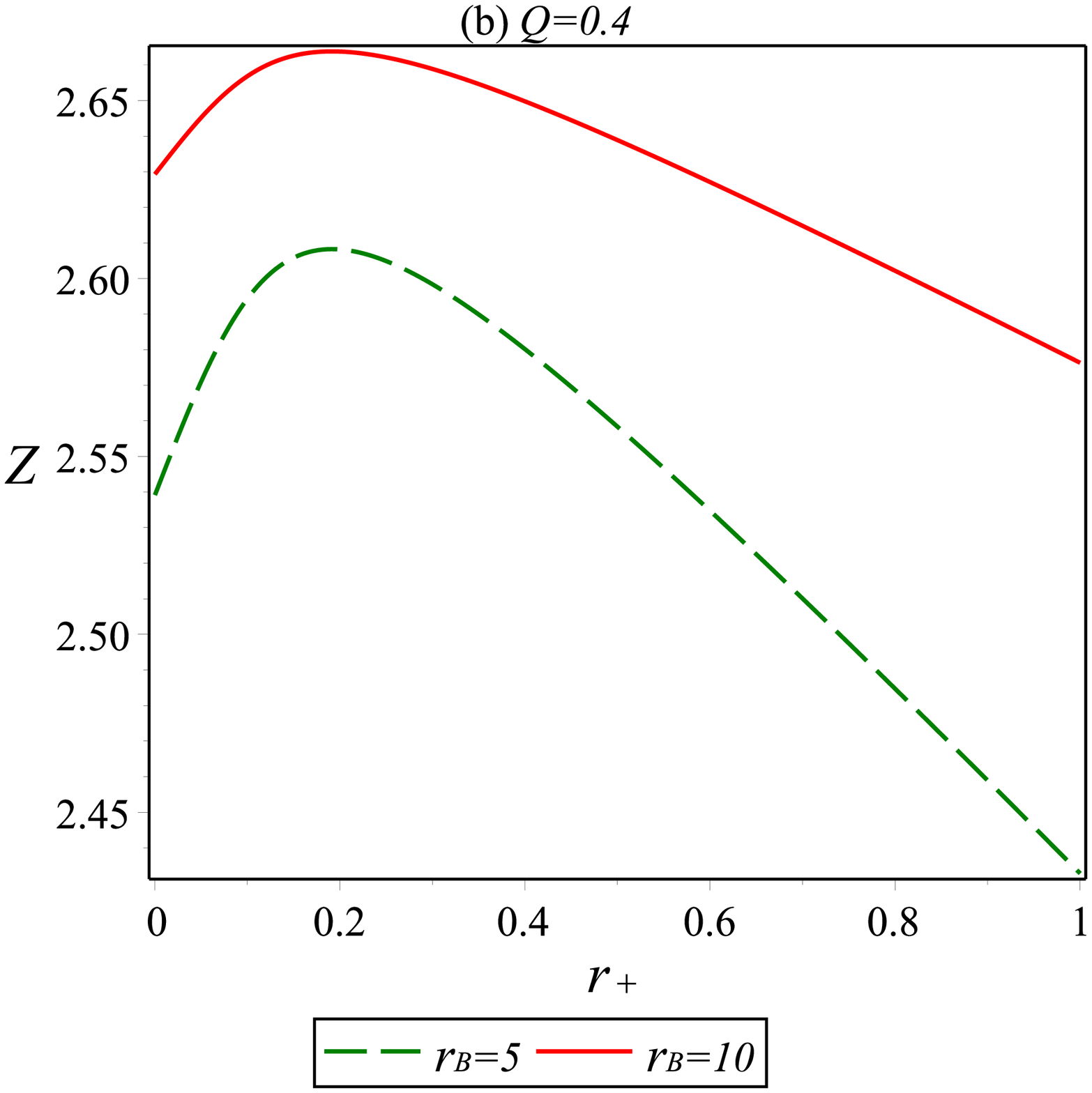}\includegraphics[width=50 mm]{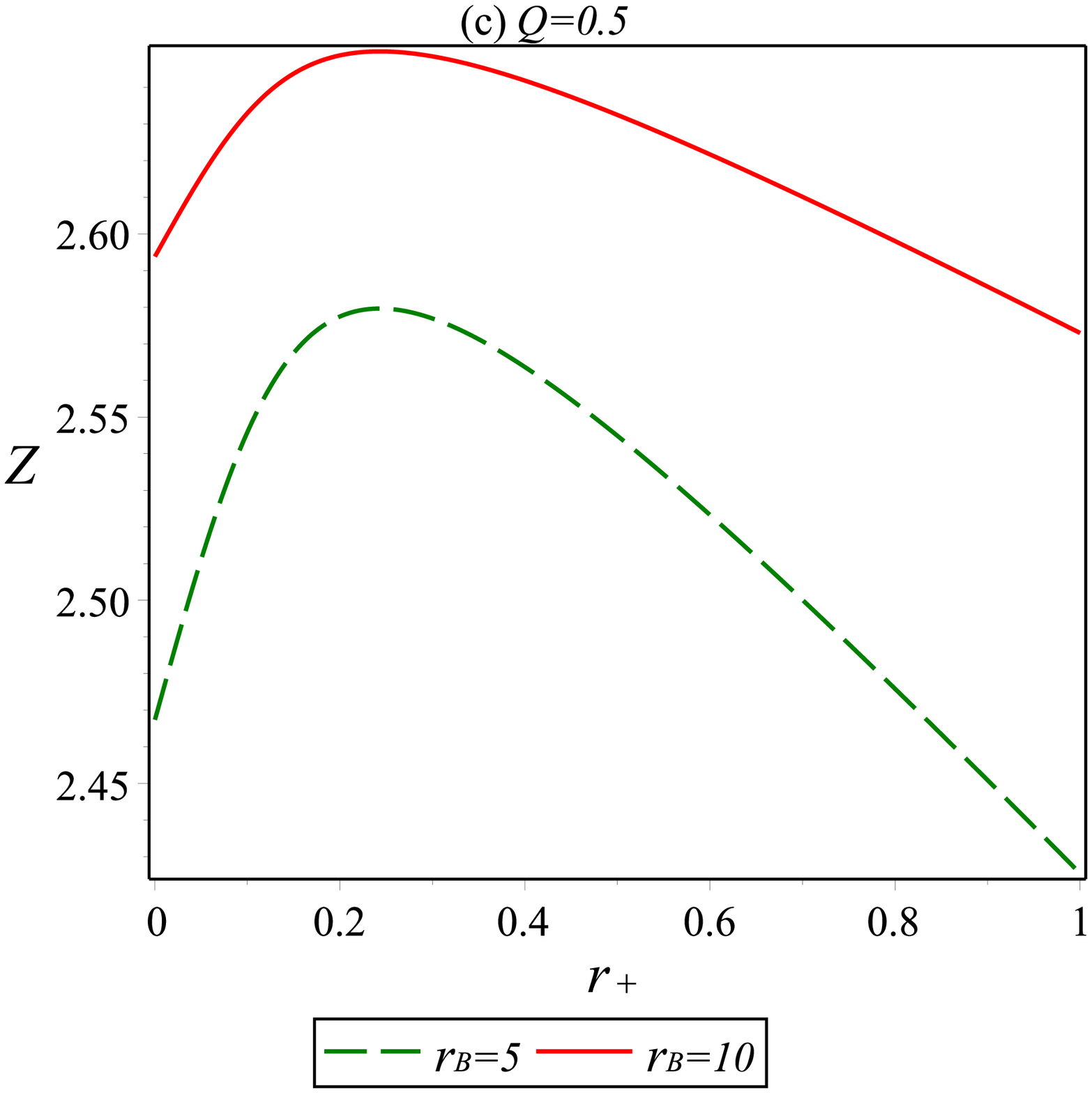}
 \end{array}$
 \end{center}
\caption{Partition function of Born-Infeld black hole in a cavity for $a=\frac{1}{300}$.}
 \label{fig6}
\end{figure}

Helmholtz free energy is related to the partition function via (in the unit of Boltzmann constant),
\begin{equation}\label{partition}
F=-T\ln{Z},
\end{equation}
where $T$ is the reciprocal temperature on the boundary of the cavity, and it is given by \cite{JHEP}
\begin{equation}\label{T}
T=\frac{T_{H}}{\sqrt{f(r_{B})}},
\end{equation}
where
\begin{equation}\label{fB}
f(r_{B})\approx1-\frac{M}{8\pi r_{B}}+\frac{Q^{2}}{4r_{B}^{2}}.
\end{equation}
Along with (\ref{F0}), (\ref{T}) and (\ref{fB}), we obtain the corrected partition function from (\ref{partition}), which can be used to obtain all the thermodynamics quantities. Behavior of the corrected partition function (\ref{partition}) is illustrated in Fig. \ref{fig6}.

\begin{figure}[h!]
 \begin{center}$
 \begin{array}{cccc}
\includegraphics[width=70 mm]{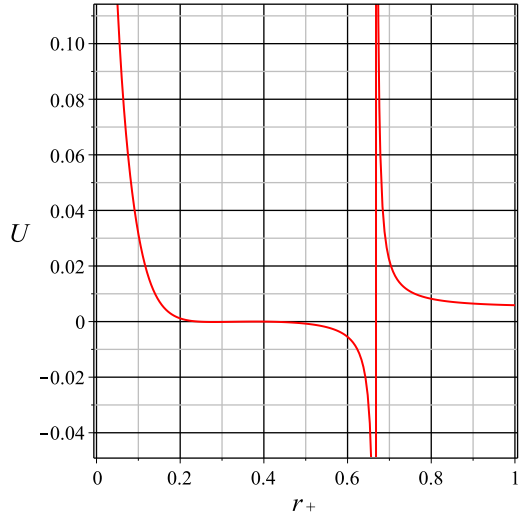}
 \end{array}$
 \end{center}
\caption{Internal energy of Born-Infeld black hole in a cavity for $a=\frac{1}{300}$ and $Q=0.4$.}
 \label{fig7}
\end{figure}

The corrected internal energy $U$ is obtained from the partition function as
\begin{equation}\label{U}
U=T^{2}\frac{d\ln{Z}}{dT}.
\end{equation}
Now we  analyze the behavior of $U$ in Fig. \ref{fig7}. Just like  the Helmholtz free energy, the internal energy $U$ is also zero at $r_{+}=r_{m}$. Here, the asymptotic region shows the first order phase transition.

\begin{figure}[h!]
 \begin{center}$
 \begin{array}{cccc}
\includegraphics[width=70 mm]{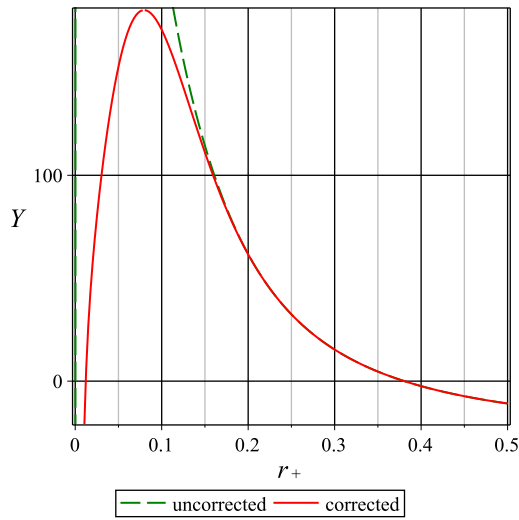}
 \end{array}$
 \end{center}
\caption{First law of thermodynamics for $a=\frac{1}{300}$, $r_{B}=10$ and $Q=0.5$.}
 \label{fig8}
\end{figure}

In the given scenario, the first law of thermodynamics is expressed as
\begin{equation}\label{first law}
Y=dU-TdS+\Phi dQ-\lambda dA,
\end{equation}
where $A=4\pi r_{B}^{2}$ is the surface area of the cavity, and $\lambda$ is a thermodynamic surface pressure defined as $(\partial U/\partial A)_{r=r_{B}}$. Also, the corresponding chemical potential is given by
\begin{equation}\label{chemical}
\Phi=\frac{4\pi{\mathcal{A}}_{t}(r_{B})}{\sqrt{f(r_{B})}},
\end{equation}
where the gauge field is given by
\begin{equation}\label{gauge}
{\mathcal{A}}_{t}(r)=\int{\frac{Qdr}{\sqrt{r^{2}+aQ^{2}}}}.
\end{equation}
The first law of thermodynamics is satisfied if $Y=0$. In Fig. \ref{fig8}, we can see that such a case comes out when $r_{+}=r_{m}$.

\section{Information  geometry}
Information geometry can be used to study the stability of black holes \cite{1111, 2222}. It can also be used to explore the points of the first-order and the second-order phase transitions. In this method, the phase transition points are identified through divergent points of the thermodynamic Ricci scalar \cite{point1, point2, point4, point5, badpa, 3dmpl}. Thus, by starting from an information metric one has to calculate the relevant Ricci scalar. The points at which the Ricci scalar diverges represent the thermodynamic phase transitions. It may be noted that it is important to use different informational theoretical metrics for analyzing the phase transition of this system. This is because it is possible that some physical information about the phase transition of black holes solutions might not be represented by specific informational theoretical metrics. It is known that all the critical points for the phase transition of a charged AdS black hole with a global monopole \cite{12qa} cannot be represented by the  Weinhold and Ruppiner metrics, and are only represented by Quevedo and Hendi-Panahiyan-Eslam-Momennia (HPEM) metrics \cite{12qb}. However, it is possible to use the Ruppiner metric to obtain important physical information about thermodynamic geometry of charged Gauss-Bonnet AdS black holes \cite{12ra}. The Weinhold metric has been used to investigate the phase transitions in a Park black hole \cite{12rw}. It has been demonstrated that for a black hole surrounded by the perfect fluid in Rastall theory, it is not possible to obtain any information from the Weinhold and Quevedo metrics, and this system can only be analyzed using the Ruppiner and HPEM metrics \cite{rh12}. In fact, more information about a black hole surrounded by the perfect fluid in the Rastall theory could be obtained from the HPEM metrics than the Ruppiner metric \cite{rh12}. Thus, different informational metrics can provide different amount of physical information about the phase transition of a black hole. So, it is important to use different information-theoretical metrics to ensure that we do not loose important physical information about critical points for the phase transition of the Born-Infeld black hole in a cavity corrected by non-perturbative corrections. Therefore, we will use several different informational theoretical metrics for analyzing it \cite{q1, q2, w1, w2, r1, r2, HPEM}.

Now we shall start with the Quevedo metrics \cite{q1, q2}. These metrics were proposed using a Legendre invariant set of metrics in the phase space, such that their pullback generates metrics on the space of equilibrium states. Thus, we will use the  two proposed Quevedo metrics ($QI$ and $QII$), which are given by
\cite{q1, q2}
\begin{equation}
ds^2= \left( SM_S+QM_Q\right) \left(-M_{SS}dS^2+M_{QQ}dQ^2\right), \hspace{2cm} (QI),
\end{equation}
and
\begin{equation}
ds^2= SM_S\left(-M_{SS}dS^2+M_{QQ}dQ^2\right), \hspace{3.5cm} (QII).
\end{equation}
It may be noted that the Ruppeiner geometry was proposed to analyze the thermodynamics fluctuations \cite{r1, r2} geometrically. The Weinhold metric was motivated from the connections between Gibbs-Duhem relation and scaling of thermodynamic potentials  \cite{w1, w2}. As they are important information metrics,  we will use both \cite{w1, w2, r1, r2} the Weinhold (W) and Ruppeiner (R) information metrics, for analyzing the Born-Infeld black hole in a cavity \begin{equation}
ds^2= M g_{ab}^{(W)}dX^adX^b,\;\;\mbox{with}\;\; g_{ab}^{(W)}=\frac{\partial^2 M }{\partial X^a \partial X^b}, \hspace{2cm} (W),
\end{equation}
and
\begin{equation}
ds^2=- \frac{M}{ M_S} g_{ab}^{(W)}dX^adX^b, \hspace{5.5cm} (R).
\end{equation}
Recently, the HPEM metric has been proposed as a new information metric \cite{HPEM}.
This was done by using a different conformal function than the Quevedo metric. The results obtained using this metric have been observed to be consistent with those obtained from the canonical ensemble \cite{HPEM1, HPEM2, HPEM3}. So, we will also use the HPEM metric for analyzing this system. The information metric of the HPEM can be expressed as \cite{HPEM}
\begin{equation}
ds^2=\frac{SM_S}{M^3_{QQ}}\left(-M_{SS}dS^2+M_{QQ}dQ^2\right), \hspace{2cm} (HPEM).
\end{equation}
In the above-mentioned metrics, $M$ is assumed to be a function of black hole charge and
entropy (we will show it shortly), and the notation $M_{XY}$ means the second-order differentiation with respect to $X$ and $Y$ variables. Now, one can calculate the Ricci scalars (${\cal{R}}$).

\begin{figure}[h!]
\begin{center}$
\begin{array}{cccc}
\includegraphics[width=70 mm]{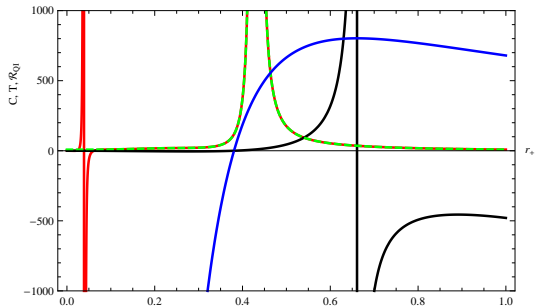}
\end{array}$
\end{center}
\caption{Ricci scalar of the Quevedo metric (${\cal{R}}_{QI}$) versus $r_{+}$ for the uncorrected (red) and corrected (green) entropy. $a=\frac{1}{300},\;Q=0.5$. Solid black line denotes specific heat, and solid blue line denotes temperature.}
\label{fig9}
\end{figure}

\begin{figure}[h!]
\begin{center}$
\begin{array}{cccc}
\includegraphics[width=70 mm]{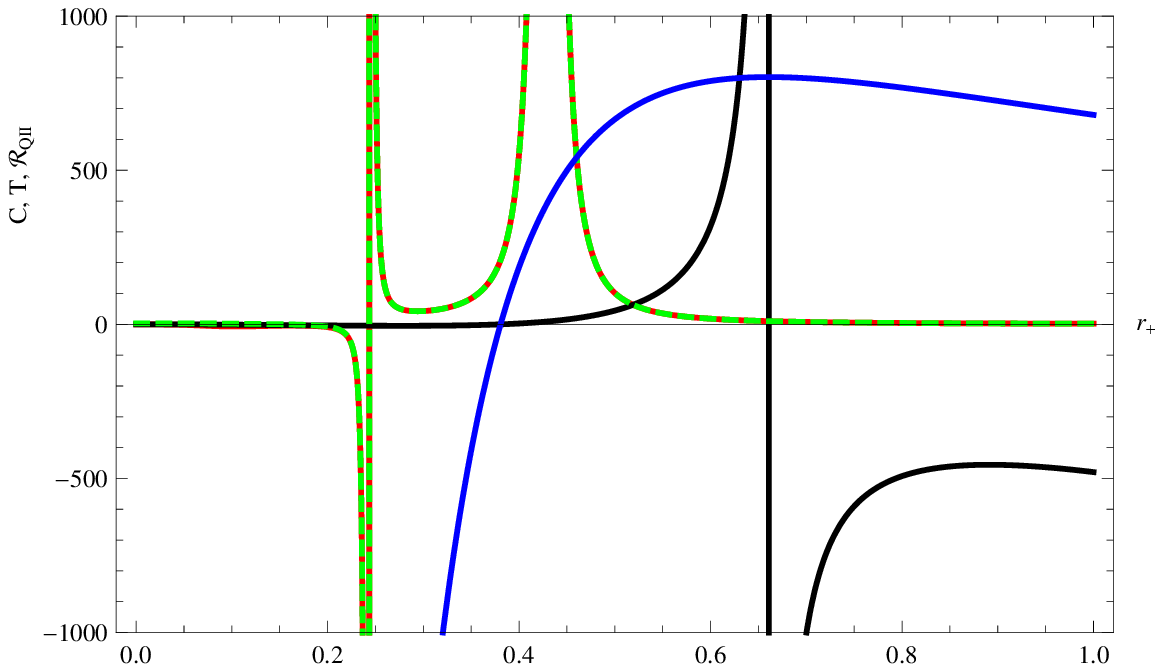}
\end{array}$
\end{center}
\caption{Ricci scalar of the Quevedo metric (${\cal{R}}_{QII}$) versus $r_{+}$ for the uncorrected (red) and corrected (green) entropy. $a=\frac{1}{300},\;Q=0.5$. Solid black line denotes specific heat, and solid blue line denotes temperature.}
\label{fig10}
\end{figure}

\begin{figure}[h!]
\begin{center}$
\begin{array}{cccc}
\includegraphics[width=70 mm]{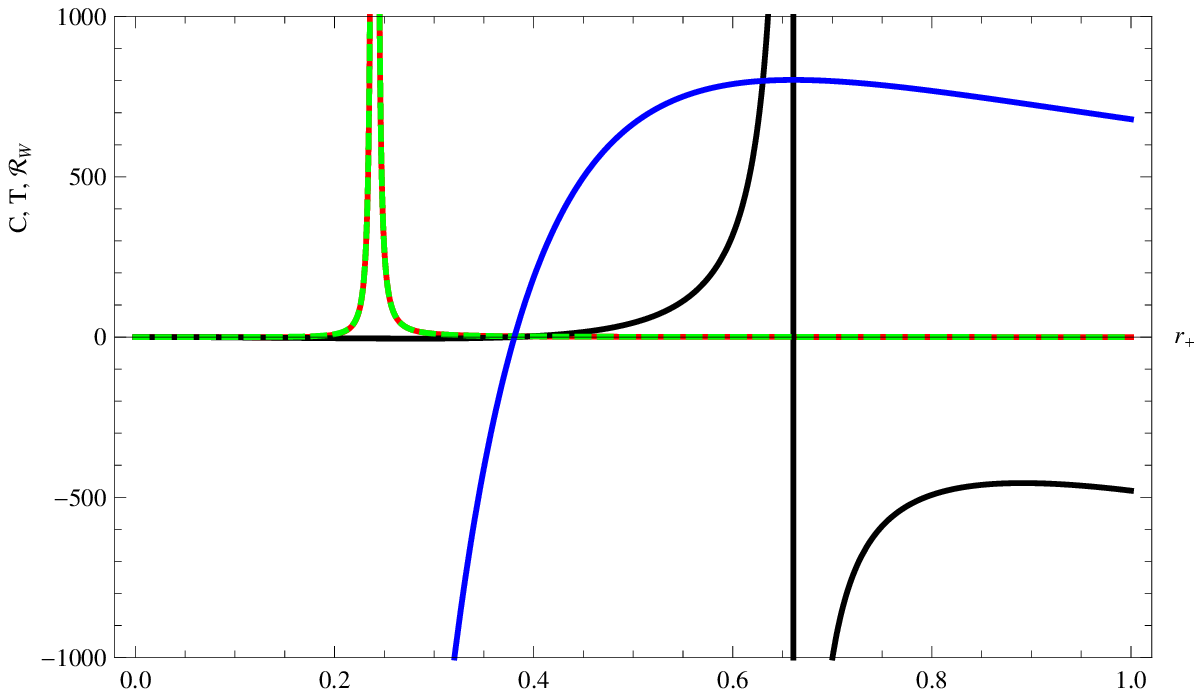}
\end{array}$
\end{center}
\caption{Ricci scalar of the Weinhold metric (${\cal{R}}_{W}$) versus $r_{+}$ for the uncorrected (red) and corrected (green) entropy. $a=\frac{1}{300},\;Q=0.5$. Solid black line denotes specific heat, and solid blue line denotes temperature.}
\label{fig11}
\end{figure}

\begin{figure}[h!]
\begin{center}$
\begin{array}{cccc}
\includegraphics[width=70 mm]{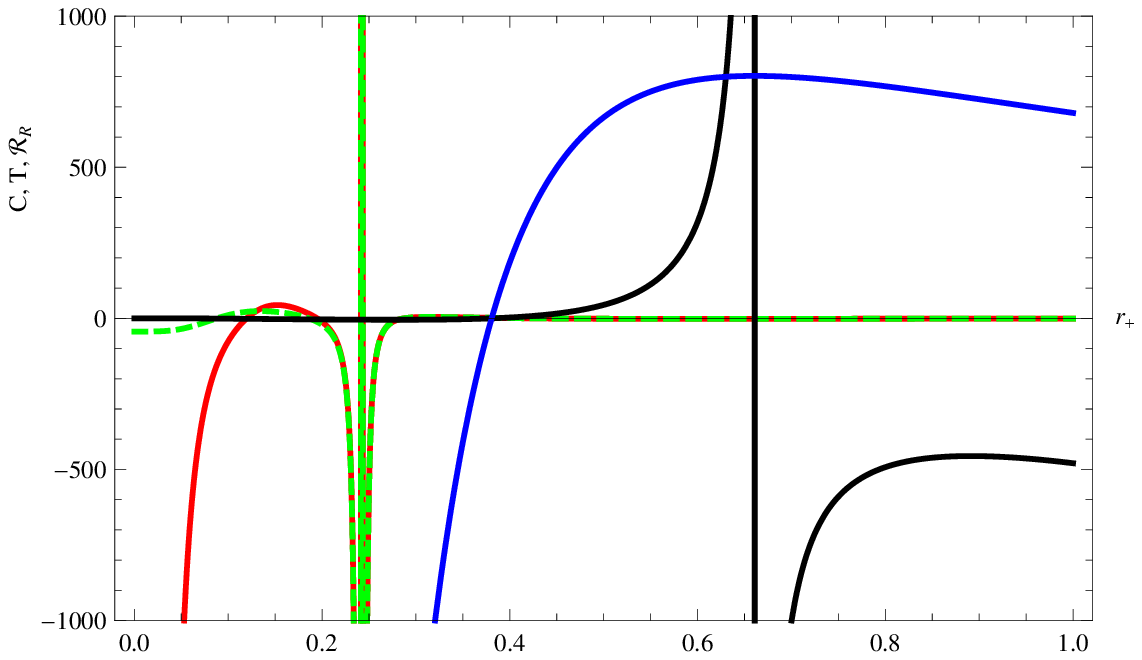}
\end{array}$
\end{center}
\caption{Ricci scalar of the Ruppeiner metric (${\cal{R}}_{R}$) versus $r_{+}$ for the uncorrected (red) and corrected (green) entropy. $a=\frac{1}{300},\;Q=0.5$. Solid black line denotes specific heat, and solid blue line denotes temperature.}
\label{fig12}
\end{figure}

\begin{figure}[h!]
\begin{center}$
\begin{array}{cccc}
\includegraphics[width=70 mm]{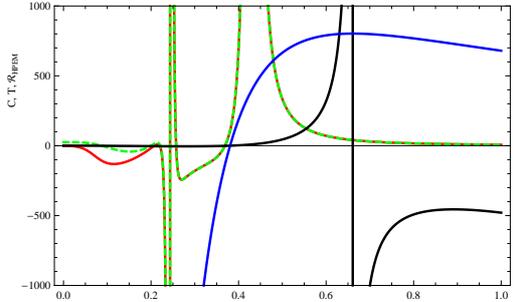}
\end{array}$
\end{center}
\caption{Ricci scalar of the HPEM metric (${\cal{R}}_{HPEM}$) versus $r_{+}$ for the uncorrected (red) and corrected (green) entropy. $a=\frac{1}{300},\;Q=0.5$. Solid black line denotes specific heat, and solid blue line denotes temperature.}
\label{fig13}
\end{figure}

Making use of Eq. (\ref{S}), one can obtain the black hole's horizon radius $r_+$ as a function of corrected entropy $S$, that is
\begin{equation}\label{hor-radi}
r_+ =\frac{\sqrt{\xi (S)}}{4 \pi},
\end{equation}
where, $\xi (S)= S+ LW(-e^{-S})$ and $ LW(x)$ is the Lambert function which satisfies the equation $ LW(x) e^{ LW(x)}=x$. For more details on Lambert function see, for example, Ref. \cite{exp}. By substituting $r_+$ from Eq. (\ref{hor-radi}) into Eq. (\ref{M}), we obtain the explicit form of the black hole mass as a function of charge and entropy. That is
\begin{equation}
M(S,Q)=2 \sqrt{\xi (S)}  \left[1- \frac{16 \pi^2Q^{2}}{6\sqrt{\xi (S) ^{2}+aQ^{2}(4\pi)^4}+6 \xi (S)  }+\frac{(4\pi)^2Q^{2}}{3 \xi (S)  }\hspace{1mm} {}_{2}F_{1}(\frac{1}{4},\frac{1}{2},\frac{5}{4};-\frac{aQ^{2}(4\pi)^2}{ \xi (S)})\right],
\end{equation}
where ${}_{2}F_{1}(x)$ is the generalized hypergeometric function.
In order to analyze the effects of non-perturbative exponential corrections on the information geometry of the black holes, we have plotted the results of calculations in Figs. \ref{fig9}-\ref{fig13}.

In the case of the Quevedo metric in the first kind (Fig. \ref{fig9}), comparing to the Fig. \ref{fig4}, we obtain the first-order phase transition for both corrected and uncorrected thermodynamics. In the case of corrected thermodynamics, we can see an asymptotic behavior. On the other hand, for the Quevedo metric of the second kind, we do not observe any important difference between corrected and uncorrected thermodynamics (see Fig. \ref{fig10}). Thus, in this case, the effects of exponential correction are negligible. The first-order phase transition also exists in this metric. Ricci scalar of the Weinhold metric represented by Fig. \ref{fig11}. Now for this metric, we do not observe any phase transition. Similar behavior is observed for the Ruppeiner metric. This is illustrated by Fig. \ref{fig12}. Finally, we plot the Ricci scalar of the HPEM metric in Fig. \ref{fig13}. The first and second-order phase transition happens for both corrected and uncorrected thermodynamics. Thus, the HPEM metric can be used to obtain important information about this system. We observe a separate curve near critical points for these cases.

\section{Conclusions}\label{sec4}
We have analyzed non-perturbative quantum corrections to a Born-Infeld black hole in a spherical cavity. In the Jacobson formalism, the quantum corrections are related to the thermal fluctuations of a black hole; and, therefore, we have used the thermal fluctuations to obtain the non-perturbative quantum corrections to this system. These corrections can be neglected at large distances, but they produce interesting modifications to the thermodynamics of this system at short distances. We have explicitly obtained the correction to the entropy of this system and observed how such corrections to the entropy of this system can produce a non-trivial modification to the thermodynamic behavior of this system. We have also analyzed the effects of this correction on various different information metrics. Thus, these thermodynamic metrics have been used to obtain non-perturbative corrections produce to the information geometry of this system.

We would like to point out that other interesting non-perturbative corrections to the thermodynamics of black holes have recently been studied. Thus, the corrections to the black hole thermodynamics from the finiteness of string theory has been studied using the T-duality of the center of mass Green's function for bosonic strings \cite{l1, t2}. This has been generalized to higher dimensions, and it has been observed that such a modification of the thermodynamics of black holes can explain the absence of mini black holes at the LHC \cite{t4}. In fact, similar arguments have also been made with the modification of black hole thermodynamics from gravity's rainbow \cite{l5} and minimal length \cite{l6}. It would be interesting to analyze such effects using the modification to the black hole thermodynamics studied here. This is expected to produce interesting consequences for the detection of mini black holes at the LHC.

\section*{Acknowledgement}
S.D. acknowledges the support of research grant (DST/INSPIRE/04/2016/001391) by DST-INSPIRE, Govt. of India.

\end{document}